\documentclass[10pt,twocolumn]{IEEEtran}

\usepackage{amsmath}
\usepackage{amsfonts}
\usepackage{epsfig}
\usepackage{amssymb}
\usepackage[nospace]{cite}
\usepackage{color,soul}
\usepackage{subfigure}
\usepackage{multirow}
\usepackage{rotating}
\usepackage{graphicx}
\usepackage{tabularx}
\usepackage{array}
\usepackage{blindtext}
\usepackage{ragged2e}
\usepackage{xcolor,colortbl}
\definecolor{Gray}{gray}{0.85}

\DeclareMathOperator*{\argmax}{arg\,max} 
\usepackage{amsthm,amssymb,amsmath,bm}
\usepackage{fancyhdr}
\usepackage[subfigure]{tocloft}
\usepackage[font={small}]{caption}
\usepackage{graphicx}
\usepackage[caption=false]{subfig}
\usepackage[utf8]{inputenc}
\usepackage[english]{babel}
\newtheorem{theorem}{Theorem}

\newtheorem{lemma}{Lemma}
\usepackage{lipsum} 
\usepackage{algorithm,algpseudocode}
\makeatletter
\newcommand{\algmargin}{\the\ALG@thistlm}
\makeatother
\newlength{\whilewidth}
\algdef{SE}[parWHILE]{parWhile}{EndparWhile}[1]
  {\parbox[t]{\dimexpr\linewidth-\algmargin}{%
     \hangindent\whilewidth\strut\algorithmicwhile\ #1\ \algorithmicdo\strut}}{\algorithmicend\ \algorithmicwhile}%
\algnewcommand{\parState}[1]{\State%
  \parbox[t]{\dimexpr\linewidth-\algmargin}{\strut #1\strut}}
\usepackage{tabularx}
\definecolor{Gray}{gray}{0.85}

\begin{document}
\vspace{-20mm}
\title{Intelligent Reflecting Surface Assisted Secret Key Generation Under Spatially Correlated Channels in Quasi-Static Environments}

\author{Vahid Shahiri, Hamid Behroozi,~\IEEEmembership{Member,~IEEE} and Ali Kuhestani, ~\IEEEmembership{Member,~IEEE}
	\thanks{V. Shahiri and H. Behroozi are with the Electrical Engineering Department, Sharif		University of Technology, Tehran, Iran. E-mail: vahid.shahiri@ee.sharif.edu, behroozi@sharif.edu}
	\thanks{A. Kuhestani is with the Electrical and Computer Engineering Department, Qom University of Technology, Qom, Iran. Email: kuhestani@qut.ac.ir}}
\maketitle
{\vspace{-10mm}}
\begin{abstract}
Physical layer key generation (PLKG) can significantly enhance the security of classic encryption schemes by enabling them to change their secret keys significantly faster and more efficient. However, due to the reliance of PLKG techniques on channel medium, reaching a high secret key rate is challenging in static environments. Recently, exploiting intelligent reflecting surface (IRS) as a means to induce randomness in static wireless channels has received significant research interest. However, the impact of spatial correlation between the IRS elements is rarely studied. To be specific, for the first time, in this contribution, we take into account a spatially correlated IRS which intends to enhance the secret key generation (SKG) rate in a static medium. Closed form analytical expressions for SKG rate are derived for the two cases of random phase shift and equal random phase shift for all the IRS elements. We also analyze the temporal correlation between the channel samples to ensure the randomness of the generated secret key sequence. We further formulate an optimization problem in which we determine the optimal portion of time within a coherence interval dedicated for the direct and indirect channel estimation. We show the accuracy and the fast convergence of our proposed sequential convex programming (SCP) based algorithm and discuss the various parameters affecting spatially correlated IRS assisted PLKG.   
\end{abstract}
\IEEEpeerreviewmaketitle
\section{Introduction}
\IEEEPARstart{W}{ireless} communications medium is intrinsically prone to the malicious eavesdropping attempts due to its broadcast nature. With the emergence of dense and widely distributed wireless networks i.e. sixth generation mobile communications (6G) and Internet of things (IoT), seeking a lightweight security approach which is able to combat the emerging modern threats has become crucial. Traditionally, the security is preserved by utilizing symmetric key cryptography (SKC) techniques such as stream ciphers, data encryption standard (DES) \cite{DES} and advanced encryption standard (AES) \cite{AES} or by using asymmetric key cryptography (AKC) methods such as Rivest–Shamir–Adleman (RSA) \cite{RSA} scheme. These higher layer security schemes had tremendous contributions in maintaining the confidentiality of communications over the previous decades. However, these encryption methods need to get boosted with physical layer security (PLS) techniques to cope with the emerging threats in modern networks.

Classic encryption schemes suffer from two major drawbacks. Specifically, the AKC methods are not preferred in networks with limited resources in their nodes i.e. in IoT. This is because AKC demands high computational resources due to it's complex mathematical operations. Thus SKC is a more desirable option for IoT networks due to it's low-complexity implementation \cite{SKC_over_AKC}. However, SKC requires the encryption keys to be distributed among the nodes before they start transferring data. This key generation process requires a complicated structure to generate and distribute the common keys between the nodes of a widely distributed network. In these cases, physical layer key generation (PLKG) techniques can be utilized to generate and distribute the random keys between the nodes \cite{frontier}. These techniques exploit the reciprocal characteristics of wireless channels as great sources of common randomness to generate random keys \cite{Zhang_1}. Another drawback is the vulnerability of classic schemes to quantum computer attacks \cite{quantum_attack}. The AKC methods rely on complex mathematical algorithms that are not scalable and thus are easily broken by a quantum computer. However, SKC schemes can be enhanced through increasing the length of the encryption keys \cite{quantum_attack}. PLKG techniques again come in handy here to help the SKC generate long random sequences of keys to boost it's strength in combating quantum computer attacks \cite{frontier}. 

Generally, the PLKG process comprises of four phases, i.e., random sharing, quantization, information reconciliation and privacy amplification \cite{Zhang_1}. During the random sharing phase, the two parties exchange pilots in time division duplex (TDD) mode to estimate the channel coefficients and exploit it as their source of common randomness. These real value coefficients are then converted to binary sequences through the quantization process \cite{quantization}. Generally the mismatches occur  during the channel estimation process of the parties. The two nodes use methods such as cosets of binary linear codes to compensate for these mismatches in information reconciliation phase \cite{information_reconciliation}. Finally, in privacy amplification phase the possible leakage of the generated keys to the eavesdroppers in the previous steps is wiped out \cite{privacy_amplification}. These four steps highlight that PLKG relies on reciprocity of physical medium characteristics to generate identical keys in two nodes and it's security is guaranteed by the inherent randomness in the physical medium.  

The randomness of the generated key is a vital requirement which is guaranteed by the temporal decorrelation between the sampled channel coefficients \cite{temporal_decorrelation}. Temporal decorrelation can be achieved in wireless networks with mobility in their nodes or their surroundings. However, this requirement is not fulfilled in static environments such as IoT networks \cite{Aldaghri}. Moreover, the level of mobility in the wireless medium may not be adequate to generate secret keys at high rate. Accordingly, \cite{Aldaghri,mimo_random_precoding,scrambling} have proposed various solutions to overcome the slow rate of PLKG in static environments. Specifically, in \cite{Aldaghri} the end users deploy random pilot constellations to induce randomness in the received signals. The authors in \cite{mimo_random_precoding}, induce randomness in multiple-input-multiple-output (MIMO) by designing random precoding vectors. In \cite{scrambling}, the correlated eavesdropper channel is scrambled through utilizing artificial noise. All of these studies focus on inducing randomness at user ends to enhance secret key generation (SKG) rate. 

Very recently, increasing the rate of channel randomness by intelligent reflecting surfaces (IRS) has received extensive research interest \cite{OTP,Quasi_static,LOS_Dominated,mmWave_IRS,random_matrix,discrete_phase,lowentropy_IRS}. Specifically, In \cite{OTP}, the authors for the first time have proposed to exploit the random shifts in IRS elements to induce randomness in the wireless channel and increase the SKR in static environments. They argue this method paves the way to the perfectly secure one-time pad (OTP) communications. In \cite{Quasi_static}, a four step protocol is designed to add randomness in a quasi-static environment. In the proposed protocol, the direct and reflective paths are estimated at each coherence time of the channel and their randomness are also exploited to enhance the SKR. In \cite{LOS_Dominated}, the authors proposed an attack model consisting of several eavesdroppers which aim to jeopardize the IRS assisted PLKG in a static line-of-sight (LOS) dominated channel. In \cite{mmWave_IRS}, the potential of IRS induced randomness in millimeter wave communications is studied. The authors in \cite{random_matrix}, have considered designing pilot signals based on random matrix theory for IRS assisted PLKG in static environments. This method avoids the leakage of generated secret keys to the eavesdropper caused due to using globally known pilot signals. In \cite{discrete_phase}, the theoretical boundaries for SKR by assuming discrete phase shifts in IRS elements are studied. Finally, the authors in \cite{lowentropy_IRS} have performed the first practical study on IRS assisted PLKG in static environments. They showed that their implemented scheme can achieve $97.39$ $\textrm{bps}$ SKR, while passing standard randomness tests. In our proposed system model, we consider the practical issue of spatial correlation present between the IRS elements. We note that, none of the above studies have considered spatial correlation in IRS in their system models.     

In another research line, the IRS is deployed to enhance the SKG rate by assisting the transceivers in conveying their signals \cite{SKG_TVT,LSP_SKG,Guyue_TIFs,arxiv_Guyue_multiantenna}. Specifically, the authors in \cite{SKG_TVT} have derived the minimum achievable SKR and developed an optimization framework for the IRS reflecting coefficients to maximized the derived secret key capacity lower bound. In \cite{LSP_SKG}, the SKR expression for the IRS assisted PLKG is deduced. The authors in \cite{Guyue_TIFs}, have considered a scenario in which a base station intends to generate secret keys with multiple user terminals (UTs) when the direct path is blocked and the signals are conveyed with an IRS. The two cases of independent and correlated channels between the channels of UTs are studied and SKR maximization frameworks have been proposed for the two cases. Moreover, in \cite{arxiv_Guyue_multiantenna}, the SKR improvement through deploying IRS when the two nodes are equipped with multiple antennas is considered. The authors have proposed an optimization algorithm to maximize SKR by designing the IRS passive beamforming. Additionally, \cite{Guyue_MWC},\cite{manipulating_attack} view the IRS as an attacker to the PLKG schemes. In \cite{Guyue_MWC}, after studying the constructive aspects of deploying IRS in PLKG i.e. in static and wave-blockage environments, the authors argue that the IRS can be utilized by an attacker to perform jamming and leakage attacks. Furthermore, the authors in \cite{manipulating_attack}, proposed an attack model in which an IRS reduces the wireless channel reciprocity by rapidly changing the RIS reflection coefficient in the uplink and downlink channel probing step. A method to detect and counteract this attack is also proposed. 

All the reviewed studies on IRS assisted PLKG assume independent reflective channels in IRS. Recently, the spatial correlation between the IRS elements is modeled in \cite{Bjornson_correlation}. The authors argue that any IRS deployed in a two-dimensional rectangular grid is subject to spatially correlated fading. This property holds for any practical IRS since it is by definition two-dimensional. The model has been widely used to consider the practical aspects of IRS deployment \cite{coverage_anastasios_1,coverage_anastasios_2,outage_IRS,performance_IRS}. In this contribution, for the first time, we investigate the impact of spatial correlation between the IRS elements in PLKG for static environments. The presence of spatial correlation in IRS elements makes our mathematical analysis of SKR challenging compared to the state-of-the-art. Furthermore, we offer an optimization framework for SKR which basically explores the intrinsic limitation in deploying the IRS to reach OTP encryption in static wireless environments. The main contributions of this paper are summarized as follows: 
\begin{itemize}
\item We consider a spatially correlated IRS which randomly changes the phases of its elements to induce artificial randomness in the static environment. This is the first time that the spatial correlation between the IRS elements is considered in such an application. 
\item We study the two cases of random phase shift for all the elements and random equal phase shift in every change of the phase of the elements and extract the SKG rate for each of these cases. It is the first time that the random equal phase shift in each phase change is considered in an IRS aided PLKG scenario for static environments.
\item Due to the presence of spatial correlation, it is not possible to directly incorporate the central limit theorem (CLT). Accordingly, we present a mathematical framework which finally leads to deriving a closed-form expression for SKG rate.
\item We derive the temporal correlation between the two channel samples for both of the above cases. We show that unlike the results presented in the literature, when the realistic IRS model is deployed, it is needed to subtract the direct channel from the indirect channel samples to generate a random secret key.
\item We propose to generate secret keys from both direct and indirect probing results and formulate an optimization problem to derive the optimum choice of dedicated time for direct and indirect probing and the optimum number of times that the IRS should change the phase of its arrays to maximize SKG rate. We propose a sequential convex programming (SCP) algorithm which is shown to be accurate and fast converging.   
\end{itemize}            

\section{System Model} \label{Sec Model}
In our proposed SKG system model, Alice and Bob as legitimate users aim to generate identical secret keys over the public channel in the presence of a passive eavesdropper (Eve). Alice, Bob and Eve are all equipped with a single antenna. Due to the relatively large coherence time in the channel between Alice and Bob, an IRS (Rose) assists them to increase the KGR. Alice and Bob probe the channel in time-devision duplex (TDD) mode to acquire correlated measurements of the shared channel to extract secret keys. Through this process, the Eve strives to obtain information on the generated secret key by listening to Alice and Bob's transmissions over the public channel. The IRS with spatial correlation between its elements acts as a trusted node which is trying to enhance the KGR by randomly shifting the phase of its elements. By doing this the IRS is able to induce virtual fast fading channel and introduce artificial randomness to the propagation environment. 
\begin{figure}[t!]
	\begin{center}
		\includegraphics[width=3.6in,height=2.5in]{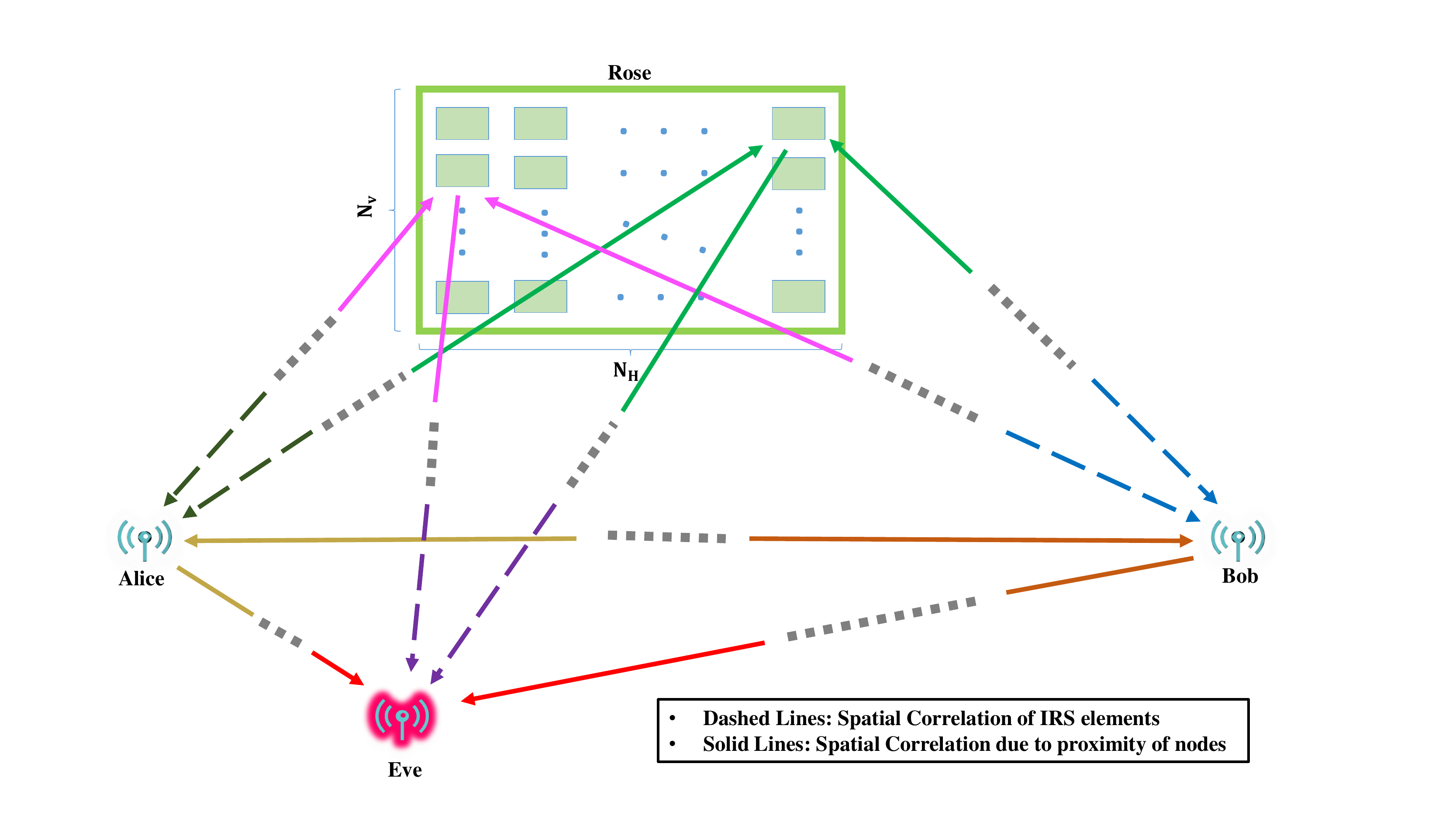}
		\caption{System model: An spatially correlated IRS aids Alice and Bob to generate secret keys from the static wireless channel in the presence of an eavesdropper.}
		\label{System_Model}
	\end{center}
\end{figure}

We assume the channels between the nodes are block-fading Rayleigh channels with coherence time $T_c$. This means that during the time interval $T_c$ the channel coefficients remain constant. As shown in Fig. X, in order to fully exploit the randomness induced by IRS, we design a two step protocol, namely the direct channel estimation and the sub-reflecting channel estimation. Specifically, we first estimate the direct channel to exploit its randomness. Moreover, these measurements will also be used in step two to mitigate the influence of the direct channel. Later in Section \ref{correlation-Sec} we will show this can lead to have negligible correlation between two consecutive probings of Alice and Bob. We assume that the time assigned for direct channel probing by each of the legitimate parties is equal to $T_d=T_{los}/2$, ($T_{los}<T_c$). 

We further assume that the IRS has $N$ reflecting elements and during step two for each channel coherence time, it changes the phase of these elements $P$ times. Accordingly, the effective coherence time of the resulting propagation medium becomes $T_e= \frac{T_r}{P}$, where $T_r$ is the time assigned for key generation from sub-reflecting channels, ($T_r<T_c$). This means that to generate identical secret keys, Alice and Bob should exchange pilot signals during the time interval $T_e$ in which the propagation environment parameters are the same for both of them. They send pilots during each time-slot which is equal to $T_s= \frac{T_e}{2}$ and take turns in sending pilot signals i.e. Alice sends pilots in even time-slots and Bob sends his pilots in odd time-slots. After all, the pilot exchange phase in step two takes $T_r=PT_e=2PT_s$ long. The remaining $T_m=T_c-T_{los}-T_r$ time is dedicated to exchange the data encrypted by using keys generated from the previous two steps. 
\subsection{Step 1: Direct Channel Estimation}
We estimate the direct channel in this step to exploit its variation from block to block in secret key generation. Additionally, the direct channel remains constant during each time slot. Thus, a strong direct channel can hinder our efforts in generating a random key sequence from sub-reflecting IRS channels. To estimate the direct channel, Alice and Bob turn off the IRS and exchange public pilots with each other. Eve also receives these pilot signals. The transmitted pilot signal by Alice is received at Bob and Eve as
\begin{equation}\label{pilot_direct_1}
\mathbf{y}_{i,1}= \sqrt{P_a}h_{ai}\mathbf{x}_d+\mathbf{n}_{i,1},\quad, i\in\{b,e\},
\end{equation}
while the pilot signal received at Alice and Eve sent by Bob is
\begin{equation}\label{pilot_direct_2}
\mathbf{y}_{i,2}= \sqrt{P_b}h_{bi}\mathbf{x}_d+\mathbf{n}_{i,2},\quad, i\in\{a,e\}.
\end{equation}
In \eqref{pilot_direct_1} and \eqref{pilot_direct_2}, $\mathbf{x}_d\in \mathbb{C}^{T_d\times 1}$ is the public pilot signal, where $T_d$ is also assumed to be the length of the pilot signal in direct channel probing. Additionally, $P_a$ and $P_b$ are the transmit power of Alice and Bob and $h_{ai}\sim\mathcal{CN}(0,\beta_{ai})$ and $h_{bi}\sim \mathcal{CN}(0,\beta_{bi})$ are the direct channel coefficients from Alice to $i$, $i=\{b,e\}$ and from Bob to $i$, $i=\{a,e\}$, respectively. Moreover, $\mathbf{n}_{i,1}\mathbf{n}_{i,2}\sim\mathcal{CN}(\mathbf{0},\sigma_i^2\mathbf{I})$ are the independent and identically distributed (i.i.d.) complex additive white Gaussian noise vectors. We assume the receivers exploit the least squares (LS) method to obtain channel state information (CSI). Accordingly, the CSI measured by Bob and Eve can be written as  
\begin{equation}\label{LS_1}
\hat{h}_{ai}=\frac{\mathbf{x}_d^H\mathbf{y}_{i,1}}{\sqrt{P_a}||\mathbf{x}_d||^2}=h_{ai}+\underbrace{\frac{1}{\sqrt{P_aT_d}}n_{i,1}}_{\hat{n}_{i,1}},\quad i\in\{b,e\},
\end{equation}
while the CSI measured by Alice and Eve is
\begin{equation}\label{LS_2}
\hat{h}_{bi}=\frac{\mathbf{x}_d^H\mathbf{y}_{i,2}}{\sqrt{P_b}||\mathbf{x}_d||^2}=h_{bi}+\underbrace{\frac{1}{\sqrt{P_bT_d}}n_{i,2}}_{\hat{n}_{i,2}},\quad i\in\{a,e\}.
\end{equation}
We note that $\hat{n}_{i,1}\sim\mathcal{CN}(0,\hat{\sigma}_{i,1}^2=\sigma_i^2/(P_aT_d))$ and $\hat{n}_{i,2}\sim\mathcal{CN}(0,\hat{\sigma}_{i,2}^2=\sigma_i^2/(P_bT_d))$ are estimation noise terms and $\hat{\sigma}_{ai}^2= \beta_{ai}+\sigma_i^2/(P_aT_d)$ and $\hat{\sigma}_{bi}^2= \beta_{bi}+\sigma_i^2/(P_bT_d)$ denote the variance of the estimated channels. Moreover, $||.||^2$ denotes the Euclidean norm of a vector.
\subsection{Step 2: Overall Sub-Reflecting Channels Estimation}
At this step, during the pilot exchange phase the transmitted pilot of Alice is received at Bob and Eve as
\begin{equation} \label{Alice_pilot}
\mathbf{y}^p_{i,1}= \sqrt{P_a}(h_{ai}+\mathbf{h}^H_{ri}\mathbf{\Phi}^p\mathbf{h}_{ar})\mathbf{x}_r+\mathbf{n}_i^{\textcolor{blue}{p}},\quad i\in\{b,e\},
\end{equation}
where $\mathbf{x}_r\in \mathbb{C}^{T_s\times 1}$ is the public pilot signal and $T_s$ is the pilot signal length. Additionally, $\mathbf{h}_{ar}\in \mathbb{C}^{N\times1}$ is the channel vector for Alice-Rose link and $\mathbf{h}_{ri}\in \mathbb{C}^{N\times1}$ is the channel from Rose to the receiver $i$. Let $\mathbf{\Phi}^p=\mathrm{diag}[e^{j\phi_{1}^p},e^{j\phi_{2}^p},...,e^{j\phi_{N}^p}]$ be the diagonal phase shift matrix of IRS with size $N$ in the $p$th round of channel probing. It is possible to configure the phase shifts by uniformly quantizing the interval $[0,2\pi)$ i.e. $\{0,\frac{2\pi}{2^B},...,\frac{(2^B-1)2\pi}{2^B}\}$, where $B$ is the number of quantization bits. In \eqref{Alice_pilot}, $\mathbf{n_i}^p\sim\mathcal{CN}(\mathbf{0},\sigma_i^2\mathbf{I})$ is the independent and identically distributed (i.i.d.) complex additive white Gaussian noise vector.

Similarly, in the subsequent time-slot Bob sends his probing sequence and the received signal at Alice and Eve is
\begin{equation}\label{Bob_pilot}
\mathbf{y}^p_{i,2}= \sqrt{P_b}(h_{bi}+\mathbf{h}^H_{ri}\mathbf{\Phi}^p\mathbf{h}_{br})\mathbf{x}_r+\mathbf{n}_i^{\textcolor{blue}{p}},\quad i\in\{a,e\}
\end{equation} 
where $\mathbf{h}_{br}\in \mathbb{C}^{N\times1}$ denotes the channel from Bob to Rose. Accordingly, the equivalent estimated channels at each node can be calculated based on observations in \eqref{Alice_pilot} and \eqref{Bob_pilot} as
\begin{subequations}\label{channel_with_direct}
\begin{align}
\hat{h}_a^p&=\frac{\mathbf{x}_r^H\mathbf{y}^p_{a,2}}{\sqrt{P_b}||\mathbf{x}_r||^2}=h_{ba}+\mathbf{h}^H_{ra}\mathbf{\Phi}^p\mathbf{h}_{br}+\underbrace{\frac{1}{\sqrt{P_bT_s}}n_a^{p}}_{\hat{n}_a^p},\label{channel_estimate_a}\\
\hat{h}_b^p&=\frac{\mathbf{x}_r^H\mathbf{y}^p_{b,1}}{\sqrt{P_a}||\mathbf{x}_r||^2}=h_{ab}+\mathbf{h}^H_{rb}\mathbf{\Phi}^p\mathbf{h}_{ar}+\underbrace{\frac{1}{\sqrt{P_aT_s}}n_b^{p}}_{\hat{n}_b^p},\label{channel_estimate_b}\\
\hat{h}_{ae}^p&=\frac{\mathbf{x}_r^H\mathbf{y}^p_{e,1}}{\sqrt{P_a}||\mathbf{x}_r||^2}=h_{ae}+\mathbf{h}^H_{re}\mathbf{\Phi}^p\mathbf{h}_{ar}+\underbrace{\frac{1}{\sqrt{P_aT_s}}n_{ae}^{p}}_{\hat{n}_{ae}^p},\label{channel_estimate_ae}\\
\hat{h}_{be}^p&=\frac{\mathbf{x}_r^H\mathbf{y}^p_{e,2}}{\sqrt{P_b}||\mathbf{x}_r||^2}=h_{be}+\mathbf{h}^H_{re}\mathbf{\Phi}^p\mathbf{h}_{br}+\underbrace{\frac{1}{\sqrt{P_bT_s}}n_{be}^{p}}_{\hat{n}_{be}^p},\label{channel_estimate_be}
\end{align}  
\end{subequations}

where $\hat{n}_a^p\sim\mathcal{CN}(0,\hat{\sigma}_a^2=\sigma_a^2/(P_bT_s))$, $\hat{n}_b^p\sim\mathcal{CN}(0,\hat{\sigma}_b^2=\sigma_b^2/(P_aT_s))$, $\hat{n}_{ae}^p\sim\mathcal{CN}(0,\hat{\sigma}_{ae}^2=\sigma_e^2/(P_aT_s))$ and $\hat{n}_{be}^p\sim\mathcal{CN}(0,\hat{\sigma}_{be}^2=\sigma_e^2/(P_bT_s))$ are the estimation error at Alice, Bob and Eve. To extract secret keys from reflecting channels, Alice and Bob subtract their measurements from the step 1 from the estimated channels in step 2 to mitigate the influence of the direct channel and Eve also follows the same steps. Accordingly, the sample according to which the secret keys are generated are
\begin{subequations}\label{channel_without_direct}
\begin{align}
h_a^p&=\hat{h}_a^p-\hat{h}_{ba}=\mathbf{h}^H_{ra}\mathbf{\Phi}^p\mathbf{h}_{br}+\underbrace{\hat{n}_a^p-\hat{n}_{a,2}}_{\hat{z}_a^p},\label{sample_1}\\
h_b^p&=\hat{h}_b^p-\hat{h}_{ab}=\mathbf{h}^H_{rb}\mathbf{\Phi}^p\mathbf{h}_{ar}+\underbrace{\hat{n}_b^p-\hat{n}_{b,1}}_{\hat{z}_b^p},\\
h_{ae}^p&=\hat{h}_{ae}^p-\hat{h}_{ae}=\mathbf{h}^H_{re}\mathbf{\Phi}^p\mathbf{h}_{ar}+\underbrace{\hat{n}_{ae}^p-\hat{n}_{e,1}}_{\hat{z}_{ae}^p},\\
h_{be}^p&=\hat{h}_{be}^p-\hat{h}_{be}=\mathbf{h}^H_{re}\mathbf{\Phi}^p\mathbf{h}_{br}+\underbrace{\hat{n}_{be}^p-\hat{n}_{e,2}}_{\hat{z}_{be}^p}.\label{sample_4}
\end{align}
\end{subequations}
We note that the subtracted noise terms in \eqref{sample_1} through \eqref{sample_4} denoting the estimation noise in step 1 and step 2, are independent random variables. Accordingly,  $\hat{z}_a^p\sim\mathcal{CN}(0,\hat{\sigma}_{z_a}^2=\sigma_a^2/(P_bT_d)+\sigma_a^2/(P_bT_s))$, $\hat{z}_b^p\sim\mathcal{CN}(0,\hat{\sigma}_{z_b}^2=\sigma_b^2/(P_aT_d)+\sigma_b^2/(P_aT_s))$, $\hat{z}_{ae}^p\sim\mathcal{CN}(0,\hat{\sigma}_{z_{ae}}^2=\sigma_e^2/(P_aT_d)+\sigma_e^2/(P_aT_s))$ and $\hat{z}_{be}^p\sim\mathcal{CN}(0,\hat{\sigma}_{z_{be}}^2=\sigma_e^2/(P_bT_d)+\sigma_e^2/(P_bT_s))$.  

We further note that we take into account for correlated Rayleigh fading channel. Mathematically, the channels are described as 
\begin{equation}\label{channels}
h_{ij}\sim\mathcal{CN}(0,\beta_{ij}) \hspace{5mm} \mathbf{h}_{ir}\sim(\mathbf{0}, \mathbf{R}_{ir}) \hspace{5mm} i,j=\{a,b,e\},
\end{equation}.

In \eqref{channels}, $\beta_{ij}$ is the path-loss between $i$ and $j$ and $\mathbf{R}_{ir}$ is the correlation matrix of the IRS elements. Here we adopt the IRS channel correlation model proposed in \cite{Bjornson_correlation}. Accordingly, $\mathbf{R}_{ir}$ is given by
\begin{equation}\label{correlation1}
\mathbf{R}_{ir}= \underbrace{\beta_{ir}d_Hd_V}_{\kappa_{ir}}\mathbf{R},
\end{equation}
in which
\begin{equation}\label{correlation2}
[\mathbf{R}]_{i,j}=\mathrm{sinc}(2||\mathbf{u}_i-\mathbf{u}_j||/\lambda) \hspace{10mm} i,j=1,...,N.
\end{equation}

In \eqref{correlation1} and \eqref{correlation2}, $d_H$ and $d_V$ are the vertical height and horizontal width of each IRS element, $\lambda$ is the wavelength of the plane wave, $\mathbf{u}_\alpha=[0,\mathrm{mod}(\alpha-1,N_H)d_H,\lfloor(\alpha-1)/N_H\rfloor d_V]^T$, $\alpha\in\{i,j\}$, where $N_H$ and $N_V$ denote the elements per row and per column of the two-dimensional rectangular IRS.

Given the channel estimates in \eqref{LS_1}, \eqref{LS_2} and \eqref{channel_without_direct}, the maximum achievable KGR is given by \cite{Maurer}
\begin{align} \label{SKG_rate}
R_{s}&=\frac{1}{2T_d}I(\hat{h}_{ab};\hat{h}_{ba}|\hat{h}_{ae},\hat{h}_{be})\nonumber\\&+\frac{1}{2PT_s}\sum_{p=1}^{P}I\bigg(h_a^p;h_b^p|h_{ae}^p,h_{be}^p\bigg).
\end{align} 
Unlike \cite{OTP}, \cite{SKG_TVT}, \cite{LSP_SKG} we are not able to incorporate central limit theorem (CLT) to calculate the second mutual information term in \eqref{SKG_rate} as we take into account for the correlated channel coefficients. To elaborate, we assume two common scenarios namely equal phase shifts and random phase shifts in IRS elements. In the following we will calculate the correlation coefficient between two different probings in the second step $\rho^{(p_1,p_2)}$ and evaluate SKG based on \eqref{SKG_rate} for the formerly mentioned scenarios.
\section{Correlation Between Subsequent Probings}\label{correlation-Sec}
The rationale behind using the IRS in our system model is to introduce randomness to a static wireless channel. Accordingly, it is vital to quantify the correlation between two consecutive probings at Alice and Bob. In this section we will evaluate this correlation for equal phase shifts (EPS) and random phase shifts (RPS) in IRS elements. Moreover, as the direct channel probing overhead may be undesirable in some use cases, here we consider the case in which the two legitimate parties generate secret keys without mitigating the influence of the direct channel. Accordingly, our analysis in this Section will include the four cases of equal phase shifts and random phase shifts in IRS elements with and without direct path in channel samples.    
\subsection{Correlation in the presence of direct path}
When the direct path is present, $\rho^{(p_1,p_2)}$ is given by the following theorem.
\begin{theorem}
The correlation coefficient between two channel samples is given by
\begin{equation}\label{correlation_coefficient}
\rho^{(p_1,p_2)}= \frac{\beta_{ij}+\mathrm{tr}\left\{\mathbf{R}_{ir}\mathbb{E}\left[{\mathbf{\Phi}^{p_2}}^H\right]\mathbf{R}_{jr}\mathbb{E}\left[\mathbf{\Phi}^{p_1}\right]\right\}}{\beta_{ij}+\mathrm{tr}\left\{\mathbb{E}\left[\mathbf{R}_{ir}{\mathbf{\Phi}^{p_l}}^H\mathbf{R}_{jr}\mathbf{\Phi}^{p_l}\right]\right\}+\hat{\sigma}_j^2},
\end{equation}  
where $l\in\{1,2\}$, $i,j\in\{a,b\}$ and $\mathrm{tr}\{.\}$  and $\mathbb{E}[.]$ denote the trace and expectation operators ,respectively. 
\end{theorem}
\textit{Proof:} Based on \eqref{channel_with_direct}, we consider two channel samples namely $s^{p_1}$ and $s^{p_2}$ as
\begin{align}
s^{p_1}= h_{ij}+\mathbf{h}^H_{rj}\mathbf{\Phi}^{p_1}\mathbf{h}_{ir}+\hat{n}_j^{p_1},\\
s^{p_2}= h_{ij}+\mathbf{h}^H_{rj}\mathbf{\Phi}^{p_2}\mathbf{h}_{ir}+\hat{n}_j^{p_2}.
\end{align}

We seek to calculate the cross correlation $\mathbb{E}[s^{p_1}s^{{p_2}^\ast}]$ and the variance $\mathbb{E}\left[|s^{p_l}|^2\right]$, $l\in\{1,2\}$. Accordingly, for the cross correlation we can write
\begin{align}
\mathbb{E}\left[s^{p_1}s^{{p_2}^\ast}\right]=&\hspace{1.5mm}\mathbb{E}\left[|h_{ij}|^2\right]+\mathbb{E}\left[\mathbf{h}^H_{rj}\mathbf{\Phi}^{p_1}\mathbf{h}_{ir}{\mathbf{h}_{ir}}^H{\mathbf{\Phi}^{p_2}}^H\mathbf{h}_{rj}\right]\nonumber\\
=&\hspace{1.5mm}\beta_{ij}+\mathbb{E}\left[\mathrm{tr}\left\{\mathbf{h}_{ir}
{\mathbf{h}_{ir}}^H{\mathbf{\Phi}^{p_2}}^H\mathbf{h}_{rj}\mathbf{h}^H_{rj}\mathbf{\Phi}^{p_1}\right\}\right]\nonumber\\
=&\hspace{1.5mm}\beta_{ij}+\mathrm{tr}\left\{\mathbf{R}_{ir}\mathbb{E}\left[{\mathbf{\Phi}^{p_2}}^H\right]\mathbf{R}_{jr}\mathbb{E}\left[\mathbf{\Phi}^{p_1}\right]\right\}.\label{corr_withdirect_fin}
\end{align}

Additionally, for the variance we have
\begin{align}
\mathbb{E}\left[|s^{p_l}|^2\right]=&\hspace{1.5mm}\mathbb{E}\left[|h_{ij}|^2\right]+\mathbb{E}\left[\mathbf{h}^H_{rj}\mathbf{\Phi}^{p_l}\mathbf{h}_{ir}
{\mathbf{h}_{ir}}^H{\mathbf{\Phi}^{p_l}}^H\mathbf{h}_{rj}\right]+\hat{\sigma}_j^2\nonumber\\
 =&\hspace{1.5mm}\beta_{ij}+\mathbb{E}\left[\mathrm{tr}\left\{\mathbf{h}^H_{rj}\mathbf{\Phi}^{p_l}\mathbf{h}_{ir}
{\mathbf{h}_{ir}}^H{\mathbf{\Phi}^{p_l}}^H\mathbf{h}_{rj}\right\}\right]+\hat{\sigma}_j^2\nonumber\\
=&\hspace{1.5mm}\beta_{ij}+\mathbb{E}\left[\mathrm{tr}\left\{\mathbf{h}_{ir}
{\mathbf{h}_{ir}}^H{\mathbf{\Phi}^{p_l}}^H\mathbf{h}_{rj}\mathbf{h}^H_{rj}\mathbf{\Phi}^{p_l}\right\}\right]+\hat{\sigma}_j^2\nonumber\\
=&\hspace{0.5mm}\beta_{ij}+\mathrm{tr}\left\{\mathbb{E}\left[\mathbf{h}_{ir}
{\mathbf{h}_{ir}}^H\right]\mathbb{E}\left[{\mathbf{\Phi}^{p_l}}^H\mathbf{h}_{rj}\mathbf{h}^H_{rj}\mathbf{\Phi}^{p_l}\right]\right\}+\hat{\sigma}_j^2\nonumber\\
=&\hspace{1.5mm}\beta_{ij}+\mathbb{E}\left[\mathrm{tr}\left\{\mathbf{h}_{rj}\mathbf{h}^H_{rj}\mathbf{\Phi}^{p_l}\mathbf{R}_{ir}{\mathbf{\Phi}^{p_l}}^H\right\}\right]+\hat{\sigma}_j^2\nonumber\\
=&\hspace{1.5mm}\beta_{ij}+\mathrm{tr}\left\{\mathbb{E}\left[\mathbf{h}_{rj}\mathbf{h}^H_{rj}\right]\mathbb{E}\left[\mathbf{\Phi}^{p_l}\mathbf{R}_{ir}{\mathbf{\Phi}^{p_l}}^H\right]\right\}+\hat{\sigma}_j^2\nonumber\\
=&\hspace{1.5mm}\beta_{ij}+\mathrm{tr}\left\{\mathbb{E}\left[\mathbf{R}_{ir}{\mathbf{\Phi}^{p_l}}^H\mathbf{R}_{jr}\mathbf{\Phi}^{p_l}\right]\right\}+\hat{\sigma}_j^2.\label{corr_withdirect_fin2}
\end{align}
Substituting into $\rho^{(p_1,p_2)}=\frac{\mathbb{E}\left[s^{p_1}s^{{p_2}\ast}\right]}{\mathbb{E}\left[|s^{p_l}|^2\right]}$, we obtain the expression in \eqref{correlation_coefficient}.
\begin{lemma}\label{corr_RPS}
The correlation coefficient between two probings when random phase shift is applied in IRS elements is given by
\begin{equation} \label{correlation_coefficient_randomshift}
\rho^{(p_1,p_2)}_{rps}= \frac{\beta_{ij}}{\beta_{ij}+\mathrm{tr}\left\{\mathbf{R}_{jr}\circ\mathbf{R}_{ir}\right\}+\hat{\sigma}_j^2},
\end{equation}
where $\circ$ denotes the Hadamard product.
\end{lemma}
\textit{Proof:} We denote the random shift phase matrix of IRS as $\mathbf{\Phi}^{p_l}= \mathrm{diag}[e^{j\phi_1^{p_l}},...,e^{j\phi_N^{p_l}}]$, where $\phi_i^{p_l}\sim\mathcal{U}(-\pi,\pi)$, $i\in\{1,...,N\}$ and $l\in\{1,2\}$. Accordingly, $\mathbb{E}\left[\mathbf{\Phi}^{p_1}\right]=\mathbb{E}\left[{\mathbf{\Phi}^{p_2}}^H\right]=\mathbf{0}$. Furthermore
\begin{align}
\mathrm{tr}&\left\{\mathbb{E}\left[\mathbf{R}_{ir}{\mathbf{\Phi}^{p_l}}^H\mathbf{R}_{jr}\mathbf{\Phi}^{p_l}\right]\right\}= \sum_{n=1}^N\sum_{m=1}^N r_{ir}^{nm}r_{jr}^{nm}\mathbb{E}\left\{e^{j(\phi_n^{p_l}-\phi_m^{p_l})}\right\}\nonumber\\
\stackrel{\text{(a)}}{=}&\sum_{n=1}^{N}r_{ir}^{nn}r_{jr}^{nn}=\mathrm{tr}\left\{\mathbf{R}_{jr}\circ\mathbf{R}_{ir}\right\},
\end{align}
where (a) holds because $\mathbb{E}\left\{e^{j(\phi_n^{p_l}-\phi_m^{p_l})}\right\}=1$, if $n=m$. Otherwise, $\mathbb{E}\left\{e^{j(\phi_n^{p_l}-\phi_m^{p_l})}\right\}=0$. Substituting in \eqref{correlation_coefficient} we obtain \eqref{correlation_coefficient_randomshift}.
\begin{lemma}\label{corr_eps}
The correlation coefficient between two probings when equal phase shift is applied in IRS elements is given by
\begin{equation}\label{correlation_coefficient_equalshift}
\rho^{(p_1,p_2)}_{eps}= \frac{\beta_{ij}}{\beta_{ij}+\mathrm{tr}\left\{\mathbf{R}_{jr}\mathbf{R}_{ir}\right\}+\hat{\sigma}_j^2},
\end{equation}
\end{lemma}
\textit{Proof:} We denote the equal phase shift in each probing as $\mathbf{\Phi}^{p_1}=e^{j\phi_{p_1}}\mathbf{I}$ and $\mathbf{\Phi}^{p_2}=e^{j\phi_{p_2}}\mathbf{I}$ where $\phi_l\sim\mathcal{U}(-\pi,\pi)$, $l\in\{1,2\}$. Accordingly, $\mathbb{E}\left[\mathbf{\Phi}^{p_1}\right]=\mathbb{E}\left[{\mathbf{\Phi}^{p_2}}^H\right]=\mathbf{0}$. Additionally 
\begin{align}
\mathrm{tr}&\left\{\mathbb{E}\left[\mathbf{R}_{ir}{\mathbf{\Phi}^{p_l}}^H\mathbf{R}_{jr}\mathbf{\Phi}^{p_l}\right]\right\}= \mathrm{tr}\left\{\mathbb{E}\left[\mathbf{R}_{ir}e^{-j\phi_{p_l}}\mathbf{R}_{jr}e^{j\phi_{p_l}}\right]\right\},\nonumber\\
=&\mathrm{tr}\left\{\mathbf{R}_{jr}\mathbf{R}_{ir}\right\}
\end{align}
Substituting in \eqref{correlation_coefficient} we obtain \eqref{correlation_coefficient_equalshift}.
\subsection{Correlation without the direct channel}
When the direct channel influence is mitigated, $\rho^{(p_1,p_2)}$ is given by the subsequent theorem. 
\begin{theorem}
The correlation coefficient between two channel samples is calculated as
\begin{equation}\label{correlation_coefficient_2}
\rho^{(p_1,p_2)}= \frac{\mathrm{tr}\left\{\mathbf{R}_{ir}\mathbb{E}\left[{\mathbf{\Phi}^{p_2}}^H\right]\mathbf{R}_{jr}\mathbb{E}\left[\mathbf{\Phi}^{p_1}\right]\right\}+\hat{\sigma}_{j,k}^2}{\mathrm{tr}\left\{\mathbb{E}\left[\mathbf{R}_{ir}{\mathbf{\Phi}^{p_l}}^H\mathbf{R}_{jr}\mathbf{\Phi}^{p_l}\right]\right\}+\hat{\sigma}_{z_j}^2},
\end{equation}
where $k=1$ if $j=b$ and $k=2$ if $j=a$. 
\end{theorem}
\textit{Proof:}Based on \eqref{channel_without_direct}, we consider two channel samples namely $s^{p_1}$ and $s^{p_2}$ as
\begin{align}
s^{p_1}= \mathbf{h}^H_{rj}\mathbf{\Phi}^{p_1}\mathbf{h}_{ir}+\hat{z}_j^{p_1},\\
s^{p_2}= \mathbf{h}^H_{rj}\mathbf{\Phi}^{p_2}\mathbf{h}_{ir}+\hat{z}_j^{p_2}.
\end{align}
We need to calculate the cross correlation between the above samples. Accordingly, we have
\begin{align}
\mathbb{E}\left[s^{p_1}s^{{p_2}^\ast}\right]=&\hspace{1.5mm}\mathbb{E}\left[\mathbf{h}^H_{rj}\mathbf{\Phi}^{p_1}\mathbf{h}_{ir}{\mathbf{h}_{ir}}^H{\mathbf{\Phi}^{p_2}}^H\mathbf{h}_{rj}\right]+\mathbb{E}\left[\hat{z}_j^{p_1}\hat{z}_j^{{p_2}^\ast}\right]\nonumber\\
\stackrel{\eqref{corr_withdirect_fin}}{=}&\hspace{1.5mm}\mathrm{tr}\left\{\mathbf{R}_{ir}\mathbb{E}\left[{\mathbf{\Phi}^{p_2}}^H\right]\mathbf{R}_{jr}\mathbb{E}\left[\mathbf{\Phi}^{p_1}\right]\right\}+\mathbb{E}\left[\hat{z}_j^{p_1}\hat{z}_j^{{p_2}^\ast}\right]\nonumber\\
\stackrel{\text{(a)}}{=}&\hspace{1.5mm}\mathrm{tr}\left\{\mathbf{R}_{ir}\mathbb{E}\left[{\mathbf{\Phi}^{p_2}}^H\right]\mathbf{R}_{jr}\mathbb{E}\left[\mathbf{\Phi}^{p_1}\right]\right\}+\hat{\sigma}_{j,k}^2,
\end{align}
where (a) holds because $\mathbb{E}\left[\hat{z}_j^{p_1}\hat{z}_j^{{p_2}^\ast}\right]=\mathbb{E}\left[\hat{n}_{j,k}\hat{n}_{j,k}^\ast\right]=\hat{\sigma}_{j,k}^2$. Moreover, 
\begin{align}
\mathbb{E}\left[|s^{p_l}|^2\right]=&\hspace{1.5mm}\mathbb{E}\left[\mathbf{h}^H_{rj}\mathbf{\Phi}^{p_l}\mathbf{h}_{ir}
{\mathbf{h}_{ir}}^H{\mathbf{\Phi}^{p_l}}^H\mathbf{h}_{rj}\right]+\mathbb{E}\left[|\hat{z}_j^{p_l}|^2\right]\nonumber\\
\stackrel{\eqref{corr_withdirect_fin2}}{=}&\hspace{1.5mm}\mathrm{tr}\left\{\mathbb{E}\left[\mathbf{R}_{ir}{\mathbf{\Phi}^{p_l}}^H\mathbf{R}_{jr}\mathbf{\Phi}^{p_l}\right]\right\}+\mathbb{E}\left[|\hat{z}_j^{p_l}|^2\right]\nonumber\\
=&\hspace{1.5mm}\mathrm{tr}\left\{\mathbb{E}\left[\mathbf{R}_{ir}{\mathbf{\Phi}^{p_l}}^H\mathbf{R}_{jr}\mathbf{\Phi}^{p_l}\right]\right\}+\hat{\sigma}_{z_j}^2.
\end{align}
. Finally, substituting into $\rho^{(p_1,p_2)}=\frac{\mathbb{E}\left[s^{p_1}s^{{p_2}\ast}\right]}{\mathbb{E}\left[|s^{p_l}|^2\right]}$, we obtain the expression in \eqref{correlation_coefficient_2}.
\begin{lemma}
The correlation coefficient between two probings when random phase shift is applied in IRS elements is given by
\begin{equation} \label{correlation_coefficient_randomshift_2}
\rho^{(p_1,p_2)}_{rps}= \frac{\hat{\sigma}_{j,k}^2}{\mathrm{tr}\left\{\mathbf{R}_{jr}\circ\mathbf{R}_{ir}\right\}+\hat{\sigma}_{z_j}^2},
\end{equation}
and for equal phase shift in IRS elements is
\begin{equation}\label{correlation_coefficient_equalshift_2}
\rho^{(p_1,p_2)}_{eps}= \frac{\hat{\sigma}_{j,k}^2}{\mathrm{tr}\left\{\mathbf{R}_{jr}\mathbf{R}_{ir}\right\}+\hat{\sigma}_{z_j}^2}.
\end{equation}
\end{lemma}
\textit{Proof:} Using the same steps in proofs of Lemmas \ref{corr_RPS} and \ref{corr_eps}, it is straightforward to obtain the expressions in \eqref{correlation_coefficient_randomshift_2} and \eqref{correlation_coefficient_equalshift_2}. 
\section{Spatially Correlated IRS Secret Key Capacity Upper Bound}
In this section we intend to calculate the mutual information terms in \eqref{SKG_rate}. Since CLT is not applicable in our system model due to the presence of spatial correlation between IRS elements, we devise a new approach which exploits eigen value decomposition (EVD) to obtain KGR. The following Theorems are prerequisites for our EVD approach. 
\begin{theorem}\label{th_gold}
Let $\Theta$ and $X$ be two independent random variables (RVs) with distributions $\Theta\sim\mathcal{U}(-\pi,\pi)$ and $X\sim\mathcal{CN}(0,\sigma^2)$, respectively. The RV, $Y=Xe^{j\Theta}$ is complex Gaussian with distribution $Y\sim\mathcal{CN}(0,\sigma^2)$.
\end{theorem}
\textit{Proof:} We denote the amplitude and phase of $X$ as $R$ and $\Psi$ where they are independent with distributions $R\sim \textit{Rayleigh}(\sigma^2/2)$ and $\Psi\sim\mathcal{U}(-\pi,\pi)$. Accordingly, we can write $Y=Re^{j(\Psi+\Theta)}=Re^{j\Phi}$. As $\Phi$ is the sum of two independent uniform Rvs distributed in the interval $(-\pi,\pi)$, it has the distribution $\Phi\sim\mathcal{U}(-\pi,\pi)$. Accordingly, $Y=Re^{j\Phi}$ has complex Gaussian distribution with mean and variance the same as $X$.   
\begin{theorem}\label{th_silver}
Let $\mathbf{\Theta}$ be an $N\times N$ random matrix defined as $\mathbf{\Theta}=\mathrm{diag}[e^{j\theta_1},...,e^{j\theta_N}]$ where $\theta_i\sim\mathcal{U}(-\pi,\pi)$, $i\in{1,...,N}$. Additionally, $\mathbf{g}$ is an $N\times1$ random vector with distribution $\mathbf{g}\sim\mathcal{CN}(\mathbf{0},\mathbf{C})$ where $\mathbf{C}^{N\times N}$ denotes the correlation matrix between the entries of $\mathbf{g}$. The random vector $\mathbf{h}=\mathbf{\Theta}\mathbf{g}$ has the distribution $\mathbf{h}\sim\mathcal{CN}\left(0,\mathrm{diag}[c_1,...,c_N]\right)$ where, $c_i, i\in{1,...,N}$ are the diagonal elements of $\mathbf{C}$. 
\end{theorem}
\textit{Proof:} For each entry of $\mathbf{h}$ we can write $h_i=g_ie^{j\theta_i}$. According to Theorem \ref{th_gold}, $h_i$ has complex Gaussian distribution with zero mean and variance $c_i$. For the correlation between two entries of $\mathbf{h}$ namely $h_i$ and $h_j$, $i\neq j$ we can write
\begin{equation}
\mathbb{E}\left\{h_ih_j^*\right\}=\mathbb{E}\left\{g_ig_j^*\right\}\mathbb{E}\left\{e^{j\theta_i}\right\}\mathbb{E}\left\{e^{-j\theta_j}\right\}=0
\end{equation} 
Accordingly, the entries of $\mathbf{h}$ are uncorrelated complex Gaussian RVs which implies they are independent with distribution $\mathbf{h}\sim\mathcal{CN}\left(0,\mathrm{diag}[c_1,...,c_N]\right)$.

In order to calculate KGR the challenging part is to deal with the correlated terms within $\mathbf{h}^H_{ri}\mathbf{\Phi}^p\mathbf{h}_{jr}$, $i,j\in\{a,b,e\}$, in \eqref{channel_without_direct}. Without the loss of generality we can express the channel gain vectors $\mathbf{h}_{ri}$ and $\mathbf{h}_{jr}$ in terms of their correlation matrices as $\mathbf{h}_{ri}=\mathbf{R}_{ri}^{\frac{1}{2}}\mathbf{g}_{ri}$ and $\mathbf{h}_{jr}=\mathbf{R}_{jr}^{\frac{1}{2}}\mathbf{g}_{jr}$, where $\mathbf{g}_{ri},\mathbf{g}_{jr}\sim\mathcal{CN}(\mathbf{0},\mathbf{I})$. Now we have the prerequisites to calculate the KGR for our system model.
\subsection{KGR for random shift in IRS elements}
To obtain KGR we firstly write the term $\mathbf{h}^H_{ri}\mathbf{\Phi}^p\mathbf{h}_{jr}$ as
\begin{equation} \label{randomshift_1}
\mathbf{h}^H_{ri}\mathbf{\Phi}^p\mathbf{h}_{jr}=\mathbf{g}_{ri}^H\mathbf{R}_{ri}^{\frac{1}{2}}\mathbf{u}_{jr},
\end{equation}
where $\mathbf{u}_{jr}=\mathbf{\Phi}^p\mathbf{h}_{jr}$. According to Theorem \ref{th_silver} and equations \eqref{correlation1} and \eqref{correlation2}, the entries of $\mathbf{u}_{jr}$ are independent identically distributed (iid) complex Gaussian RVs with distribution $\mathbf{u}_{jr}\sim\mathcal{CN}(\mathbf{o},\kappa_{jr}\mathbf{I})$. We perform the EVD on $\mathbf{R}_{ri}^{\frac{1}{2}}$ as $\mathbf{R}_{ri}^{\frac{1}{2}}=\mathbf{Q}_{ri}\mathbf{\Sigma}_{ri}\mathbf{Q}_{ri}^H$ where $\mathbf{Q}_{ri}$ is $N\times N$ complex unitary matrix and $\mathbf{\Sigma}_{ri}$ is an $N\times N$ diagonal matrix, i.e., $\mathbf{\Sigma}_{ri}=\mathrm{diag}[\lambda_{ri}^1,...,\lambda_{ri}^N]$. Accordingly we can rewrite \eqref{randomshift_1} as
\begin{equation}\label{CLT_1}
\mathbf{h}^H_{ri}\mathbf{\Phi}^p\mathbf{h}_{jr}=\underbrace{\mathbf{g}_{ri}^H\mathbf{Q}_{ri}}_\text{$\mathbf{w}_{ri}^H$}\mathbf{\Sigma}_{ri}\underbrace{\mathbf{Q}_{ri}^H\mathbf{u}_{jr}}_\text{$\mathbf{w}_{jr}$}.
\end{equation}   
Since $\mathbf{Q}_{ri}$ is a unitary matrix, the new channel vectors are $\mathbf{w}_{ri}\sim\mathcal{CN}(\mathbf{0},\mathbf{I})$, $\mathbf{w}_{jr}\sim\mathcal{CN}(\mathbf{0},\kappa_{jr}\mathbf{I})$. Now that we are dealing with the sum of independent RVs in \eqref{CLT_1}, we can assert that for $N>>1$, $\mathbf{h}^H_{ri}\mathbf{\Phi}^p\mathbf{h}_{jr}$ is complex Gaussian with $\mathcal{CN}(0,\kappa_{jr}\sum_{n=1}^{N}{\lambda_{ri}^n}^2)$. Accordingly, we could transform correlated channel vector entries to iid entries through applying Theorem \ref{th_silver} and performing EVD. 
\subsection{KGR for equal shift in IRS elements}
In the equal phase shift case for the IRS shift matrix we have $\mathbf{\Phi}^p=e^{j\phi_p}\mathbf{I}$, $\phi_p\sim\mathcal{U}(-\pi,\pi)$. Accordingly we can write
\begin{equation}
\mathbf{h}^H_{ri}\mathbf{\Phi}^p\mathbf{h}_{jr}=e^{j\phi_p}\mathbf{g}_{ri}^H\underbrace{\mathbf{R}_{ri}^{\frac{1}{2}}\mathbf{R}_{jr}^{\frac{1}{2}}}_\text{$\mathbf{\Psi}_{ij}$}\mathbf{g}_{jr}.
\end{equation}
Performing EVD on $\mathbf{\Psi}_{ij}$ we can write $\mathbf{\Psi}_{ij}=\mathbf{P}_{ij}\mathbf{\Xi}_{ij}\mathbf{P}_{ij}^H$, where $\mathbf{P}_{ij}$ is an $N\times N$ complex unitary matrix and $\mathbf{\Xi}_{ij}$ is an $N\times N$ diagonal matrix, i.e., $\mathbf{\Xi}_{ij}=\mathrm{diag}[\rho_{ij}^1,...,\rho_{ij}^N]$.
\begin{equation}
\mathbf{h}^H_{ri}\mathbf{\Phi}^p\mathbf{h}_{jr}=e^{j\phi_p}\underbrace{\mathbf{g}_{ri}^H\mathbf{P}_{ij}}_\text{$\mathbf{v}_{ri}^H$}\mathbf{\Xi}_{ij}\underbrace{\mathbf{P}_{ij}^H\mathbf{g}_{jr}}_\text{$\mathbf{v}_{jr}$}.
\end{equation}
Since $\mathbf{P}_{ij}$ is a unitary matrix, the new channel vectors are $\mathbf{v}_{ri}\sim\mathcal{CN}(\mathbf{0},\mathbf{I})$, $\mathbf{v}_{jr}\sim\mathcal{CN}(\mathbf{0},\mathbf{I})$. For $N>>1$ we can apply CLT on $x_{ij}=\mathbf{v}_{ri}^H\mathbf{\Xi}_{ij}\mathbf{v}_{jr}$ as $x_{ij}$ is the sum of independent RVs which leads to $x_{ij}\sim\mathcal{CN}(0,\sum_{n=1}^{N}{\rho_{ij}^n}^2)$. Finally,
\begin{equation}
\mathbf{h}^H_{ri}\mathbf{\Phi}^p\mathbf{h}_{jr}=e^{j\phi_p}x_{ij}=y_{ij},
\end{equation}  
in which according to Theorem \ref{th_gold}, $y_{ij}$ is a complex Gaussian RV with the same mean and variance as $x_{ij}$, $y_{ij}\sim\mathcal{CN}(0,\sum_{n=1}^{N}{\rho_{ij}^n}^2)$. 

We remark that the variance of $\mathbf{h}^H_{ri}\mathbf{\Phi}^p\mathbf{h}_{jr}$ obtained in this Section is equal to the expression calculated in the previous Section. In other words we have
\begin{align}
\kappa_{jr}\sum_{n=1}^{N}{\lambda_{ri}^n}^2=&\hspace{1.5mm}\mathrm{tr}\left\{\mathbf{R}_{jr}\circ\mathbf{R}_{ir}\right\},\\
\sum_{n=1}^{N}{\rho_{ij}^n}^2=&\hspace{1.5mm}\mathrm{tr}\left\{\mathbf{R}_{jr}\mathbf{R}_{ir}\right\}.
\end{align}
Using the well-known properties of eigen values in linear algebra, verifying the above equations is straightforward. Now that we have the statistical properties of our channel samples, we can evaluate the SKG rate.
\begin{theorem}
The maximum achievable SKG rate in IRS assisted wireless network with spatial correlation between IRS elements is   
\begin{equation}\label{total_SKG_rate}
R_{s}= \frac{1}{2T_d}\log_2\Lambda(\mathbf{\Omega})
+\frac{1}{2T_s}\log_2\Lambda(\mathbf{\Delta})
\end{equation}
\begin{figure*}
\begin{equation}
\mathbf{\Lambda}(\mathbf{v})=\frac{\left[x_1(x_3x_4-y_2^2)+2y_2y_3y_4-y_4^2x_3-y_3^2x_4\right]\left[x_2(x_3x_4-y_2^2)+2y_2y_3y_4-y_4^2x_3-y_3^2x_4\right]}{(x_3x_4-y_2^2)\left[(x_1+x_2-2y_1)(2y_2y_3y_4-y_4^2x_3-y_3^2x_4)-(x_3x_4-y_2^2)(y_1^2-x_1x_2)\right]}
\end{equation}
\hrulefill
\end{figure*} 
where by defining $\mathbf{v}=(x_1,x_2,x_3,x_4,y_1,y_2,y_3,y_4)$, $\Lambda(\mathbf{v})$ is defined at the top of the next page and  $\mathbf{\Omega}=(\Omega_1,\Omega_2,\Omega_3,\Omega_4,\omega_1,\omega_2,\omega_3,\omega_4)$, $\mathbf{\Delta}=(\Delta_1,\Delta_2,\Delta_3,\Delta_4,\delta_1,\delta_2,\delta_3,\delta_4)$. Moreover we have
\begin{align}
\Omega_1=&\hspace{1.5mm}\upsilon_{aa}=\sigma_{ab}^2+\hat{\sigma}_{a,2}^2,\\
\Omega_2=&\hspace{1.5mm}\upsilon_{bb}=\sigma_{ab}^2+\hat{\sigma}_{b,1}^2,\\
\Omega_3=&\hspace{1.5mm}\upsilon_{aeae}=\sigma_{ae}^2+\hat{\sigma}_{e,1}^2,\\
\Omega_4=&\hspace{1.5mm}\upsilon_{bebe}=\sigma_{be}^2+\hat{\sigma}_{e,2}^2,\\
\omega_1=&\hspace{1.5mm}\upsilon_{ab}=\upsilon_{ba}=\sigma_{ab}^2,\\
\omega_2=&\hspace{1.5mm}\upsilon_{aebe}=\upsilon_{beae}=\rho_{ab}\sigma_{ae}\sigma_{be},\\
\omega_3=&\hspace{1.5mm}\upsilon_{aae}=\upsilon_{aea}=\upsilon_{bae}=\upsilon_{aeb}=\rho_{be}\sigma_{ae}\sigma_{ab},\\
\omega_4=&\hspace{1.5mm}\upsilon_{abe}=\upsilon_{bbe}=\upsilon_{bea}=\upsilon_{beb}=\rho_{ae}\sigma_{ab}\sigma_{be},\\
\Delta_1=&\eta_{aa}=\mathrm{tr}\left\{\mathbf{R}_{ar}\odot\mathbf{R}_{br}\right\}+\hat{\sigma}_{z_a}^2,\\
\Delta_2=&\eta_{bb}=\mathrm{tr}\left\{\mathbf{R}_{ar}\odot\mathbf{R}_{br}\right\}+\hat{\sigma}_{z_b}^2,\\
\Delta_3=&\eta_{aeae}=\mathrm{tr}\left\{\mathbf{R}_{ar}\odot\mathbf{R}_{er}\right\}+\hat{\sigma}_{z_{ae}}^2,\\
\Delta_4=&\eta_{bebe}=\mathrm{tr}\left\{\mathbf{R}_{br}\odot\mathbf{R}_{er}\right\}+\hat{\sigma}_{z_{be}}^2,\\
\delta_1=&\eta_{ab}=\eta_{ba}=\mathrm{tr}\left\{\mathbf{R}_{ar}\odot\mathbf{R}_{br}\right\},\\
\delta_2=&\eta_{aebe}=\eta_{beae}=\rho_{ab}\sqrt{\kappa_{ar}}\sqrt{\kappa_{br}}\kappa_{re}\nonumber\\
&\times\mathrm{tr}\left\{\mathbf{R}\odot\mathbf{R}\right\},\\
\delta_3=&\eta_{aae}=\eta_{aea}=\eta_{bae}=\eta_{aeb}=\rho_{be}\sqrt{\kappa_{er}}\nonumber\\
&\times\sqrt{\kappa_{br}}\kappa_{ra}\mathrm{tr}\left\{\mathbf{R}\odot\mathbf{R}\right\},\\
\delta_4=&\eta_{abe}=\eta_{bbe}=\eta_{bea}=\eta_{beb}=\rho_{ae}\sqrt{\kappa_{er}}\nonumber\\
&\times\sqrt{\kappa_{ar}}\kappa_{rb}\mathrm{tr}\left\{\mathbf{R}\odot\mathbf{R}\right\},
\end{align}
where we have accounted for the reciprocal channels between the nodes $\sigma_{ij}^2=\beta_{ij}=\beta_{ji}, i,j\in\{a,b,e\}$. Additionally, $\rho_{ij}=J_0(2\pi d_{ij}/\lambda)$ denotes the spatial correlation coefficient between the channels of nodes $i$ and $j$, where $J_0(.)$ is the zeroth-order Bessel function of the first kind, $\lambda$ is the wave length and $d_{ij}$ is the distance between the two nodes. Moreover, $\upsilon$ and $\eta$ denote the cross-correlation between the channel samples of Alice, Bob and Eve in direct and IRS-assisted probings, respectively. Finally, $\odot$ denotes the matrix multiplication when EPS is deployed in IRS while it denotes Hadamard product in RPS mode.   
\end{theorem} 
\textit{Proof:} We consider the mutual information term associated with the indirect path in \eqref{SKG_rate} as the calculations for the direct path term is similar and straightforward. Since we showed that $\mathbf{h}^H_{ri}\mathbf{\Phi}^p\mathbf{h}_{jr}$, $i,j\in\{a,b,e\}$ terms in \eqref{channel_without_direct} have complex normal distribution for $N>>1$, the conditional mutual information can be calculated as 
\begin{align}\label{det}
I\bigg(h_a^p;h_b^p|& h_{ae}^p,h_{be}^p\bigg)= H(h_a^p,h_{ae}^p,h_{be}^p)+H(h_b^p,h_{ae}^p,h_{be}^p)\nonumber\\
&-H(h_a^p,h_b^p,h_{ae}^p,h_{be}^p)-H(h_{ae}^p,h_{be}^p)\nonumber\\
&\hspace{13mm}=\log_2\frac{\mathrm{det}(\mathbf{M}_{aaebe})\hspace{1mm}\mathrm{det}(\mathbf{M}_{baebe})}{\mathrm{det}(\mathbf{M}_{aebe})\hspace{1mm}\mathrm{det}(\mathbf{M}_{abaebe})},
\end{align}     
where $\mathrm{det}(.)$ is the matrix determinant, and 
\begin{align}\label{mat_corr}
\mathbf{M}_{abaebe}&=\mathbb{E}\left[
\begin{pmatrix}
h_a^p\\
h_b^p\\
h_{ae}^p\\
h_{be}^p
\end{pmatrix}
\begin{pmatrix}
h_a^{p^\ast}& h_b^{p^\ast}& h_{ae}^{p^\ast}& h_{be}^{p^\ast}
\end{pmatrix}
\right]\nonumber\\
&=\begin{bmatrix}
\eta_{aa} & \eta_{ab} & \eta_{aae} & \eta_{abe}\\
\eta_{ba} & \eta_{bb} & \eta_{bae} & \eta_{bbe}\\
\eta_{aea} & \eta_{aeb} & \eta_{aeae} & \eta_{aebe}\\
\eta_{bea} & \eta_{beb} & \eta_{beae} & \eta_{bebe}
\end{bmatrix}
,
\end{align}
where $\eta_{mn}=\mathbb{E}\left[h_{m}^p h_{n}^{p^\ast}\right]$, $m,n\in\{a,b,ae,be\}$ is the correlation function. Obtaining the expressions for $\eta_{aa},\eta_{bb},\eta_{aeae},\eta_{bebe},\eta_{ab}=\eta_{ba}$ is straightforward by following the steps in obtaining $\mathbb{E}\left[|s^{p_l}|^2\right]$ in Section \ref{correlation-Sec}. For the other terms i.e. $\eta_{aebe}$, we have
\begin{align}
\eta_{aebe}=&\hspace{1mm} \mathbb{E}\left[h_{ae}^ph_{be}^{p^\ast}\right]=\mathbb{E}\left[\mathbf{h}^H_{re}\mathbf{\Phi}^p\mathbf{h}_{ar}  \mathbf{h}^H_{br}\mathbf{\Phi}^{p^H}\mathbf{h}_{re}\right]\nonumber\\
=&\hspace{1mm} \mathrm{tr}\left\{\mathbb{E}\left[\mathbf{\Phi}^p\mathbf{h}_{ar}  \mathbf{h}^H_{br}\mathbf{\Phi}^{p^H}\right]\mathbb{E}\left[\mathbf{h}_{re}\mathbf{h}^H_{re}\right]\right\}\nonumber\\
=&\hspace{1mm}\mathbb{E}\left[\mathrm{tr}\left\{\mathbf{h}_{ar}  \mathbf{h}^H_{br}\mathbf{\Phi}^{p^H}\mathbf{R}_{re}\mathbf{\Phi}^p\right\}\right]\nonumber\\
=&\hspace{1mm}\mathrm{tr}\left\{\mathbb{E}\left[\mathbf{h}_{ar}  \mathbf{h}^H_{br}\right]\mathbb{E}\left[\mathbf{\Phi}^{p^H}\mathbf{R}_{re}\mathbf{\Phi}^p\right]\right\}\nonumber\\
=&\hspace{1mm}\mathrm{tr}\left\{\mathbf{R}_{ar}^{\frac{1}{2}}\mathbb{E}\left[\mathbf{g}_{ar}\mathbf{g}_{br}^H\right]\mathbf{R}_{br}^{\frac{1}{2}^H}\mathbb{E}\left[\mathbf{\Phi}^{p^H}\mathbf{R}_{re}\mathbf{\Phi}^p\right]\right\}\nonumber\\
\stackrel{\text{(a)}}{=}&\hspace{1mm}\rho_{ab}\mathrm{tr}\left\{\mathbb{E}\left[\mathbf{R}_{ar}^{\frac{1}{2}}\mathbf{R}_{br}^{\frac{1}{2}}\mathbf{\Phi}^{p^H}\mathbf{R}_{re}\mathbf{\Phi}^p\right]\right\}\nonumber\\
\stackrel{\text{(b)}}{=}&\hspace{1mm}\rho_{ab}\sqrt{\kappa_{ar}}\sqrt{\kappa_{br}}\kappa_{re}\mathrm{tr}\left\{\mathbf{R}\odot\mathbf{R}\right\},
\end{align} 
where (a) holds because $\mathbb{E}\left[\mathbf{g}_{ar}\mathbf{g}_{br}^H\right]=\rho_{ab}\mathbf{I}$ and (b) is deduced based on the proofs of lemmas \ref{corr_RPS} and \ref{corr_eps}. Following the same steps, other correlation terms in \eqref{mat_corr} can be obtained. Similarly, the determinants of the other matrices in \eqref{det} are calculated.   
\vspace{5mm}
\section{optimization of sampling period and probing time}
\begin{figure}
	\begin{center}
		\includegraphics[width=3.2in,height=3.5in]{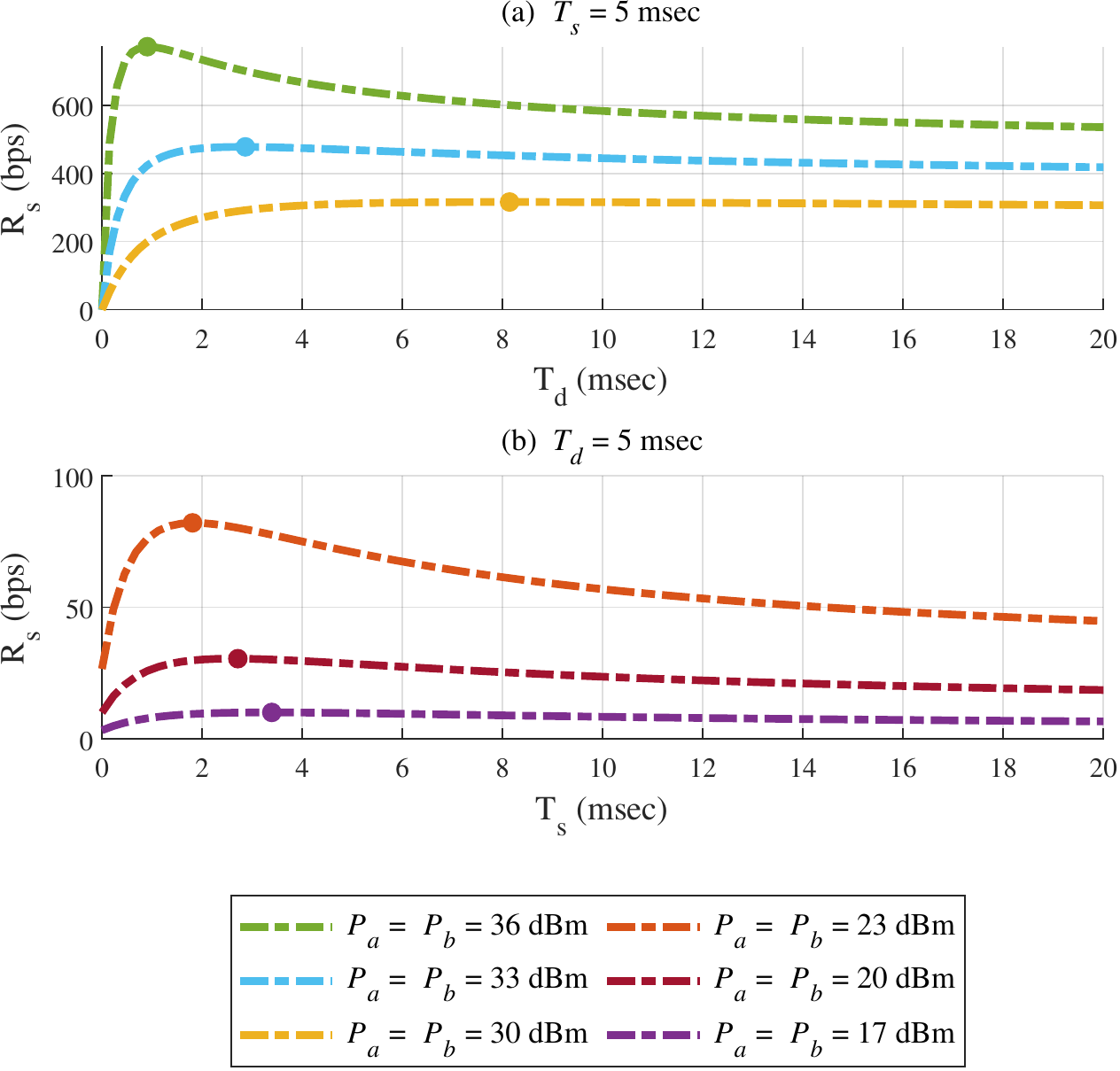} \vspace{0.0cm}
		\caption{SKG rate versus allocated time for direct probing and indirect IRS aided probing for different levels of transmit power in the nodes. }
		\label{opt_R_SKG_overview_1}
	\end{center}
\end{figure}

In this section, we intend to develop an optimization framework to determine the optimum sampling period and probing time for direct and IRS channels. In fact, the idea of implementing random shifts in IRS to enhance SKR has it's own limitations. Specifically, one may assume that by reducing the sampling period $T_s$ we can obtain more channel samples and increase the SKR. However, reducing $T_s$ will lead to the reduction of the number of bits extracted from each sample \cite{discrete_phase}. This is because reducing $T_s$ will enhance the channel estimation noise power and accordingly limit the SKR. Moreover, choosing a large $T_s$ will hinder the maximum utilization of IRS in generating secret keys. Therefore, the optimum selection of $T_s$ is vital to achieve the maximum SKR. We further note that it is important to optimally allocate the overall probing time between the direct and reflective paths as this can affect both the correlation between the samples and the maximum achievable SKR. Accordingly we formulate our optimization problem as 
\begin{subequations}\label{Opt_1}
	\begin{align}
	&{\mathop {\max }\limits_{T_d,T_r,T_s}  \; R_{s},}\label{opt_problem}\\&
	{{\rm{s}}.{\rm{t}}.: 2T_d+T_r=T_p,} \label{limit_1}\\&
	\hspace{.8cm}T_s \le \frac{T_r}{2}, \label{limit_2}\\&
	\hspace{.8cm}\max\{\rho^{(p_1,p_2)}_{s_a},\rho^{(p_1,p_2)}_{s_b}\}\le\rho^t, \label{limit_3}
	\end{align}
\end{subequations}
where $s\in\{rps,eps\}$ and $\rho^t$ and $T_p$ denote the maximum permissible correlation between the channel samples and the dedicated time for probing within a coherence time of the channel, respectively. Substituting \eqref{limit_1} into \eqref{limit_2}, we can reformulate the optimization problem as 
\begin{subequations}\label{Opt_2}
	\begin{align}
	&{\mathop {\max }\limits_{T_d,T_s}  \; R_{s},}\label{opt_problem_new}\\&
	{{\rm{s}}.{\rm{t}}.: T_s+T_d\le\frac{T_p}{2},} \label{limit_2_1}\\&
	\hspace{.8cm}\rho^{(p_1,p_2)}_{s_a}\le\rho^t, \label{limit_2_2}\\&
	\hspace{.8cm}\rho^{(p_1,p_2)}_{s_b}\le\rho^t. \label{limit_2_3}
	\end{align}
\end{subequations}
To  provide more insights on the objective function, we have plotted the secret key generation rate versus $T_d$ and $T_s$ in Fig. \ref{opt_R_SKG_overview_1} for different levels of transmit power in Alice and Bob. In both Fig. \ref{opt_R_SKG_overview_1}.a and Fig. \ref{opt_R_SKG_overview_1}.b, we can observe that there is an optimum value for $T_d$ and $T_s$, respectively which maximizes the SKG rate given in \eqref{opt_problem_new}. This observation is our motivation in raising the optimization problem in \eqref{Opt_2}.
  
\textit{Remark 1:} In order to enhance the randomness in generated secret keys, we subtracted the direct channel coefficient from the channel measurements in indirect probing. This lead to the channel estimation error (CEE) in the second term of \eqref{total_SKG_rate} to be the aggregate of CEE in both direct and indirect probings. However, maximizing $T_d$ to minimize the estimation error in the direct probing, is not an optimum strategy. This is because of the contribution of the direct channel in SKG rate, reflected in the first term of \eqref{total_SKG_rate}. Accordingly, reckless maximizing of $T_d$ can degrade the SKG rate generated by the direct path. This trade off shows the existence of an optimum value for $T_d$ which is reflected in curves of Fig. \ref{opt_R_SKG_overview_1}.a.   

\textit{Remark 2:} As stated before, reducing $T_s$ can hinder the SKG process by enhancing the CEE in the indirect probing phase. According to Fig. \ref{opt_R_SKG_overview_1}.b in the higher levels of transmit power, the optimal value for $T_s$ gets smaller. This means that in a given $T_r$, the IRS can change the phase of its arrays more rapidly in high transmit powers and accordingly, enhance the SKG rate. We will further discuss the details in the next section.   

Here, we intend to develop an optimization algorithm based on sequential convex programming (SCP), to derive the optimum $T_s$ and $T_d$ namely, ${T_s}^\star$ and ${T_d}^\star$, for \eqref{Opt_2}. We remark that the objective function \eqref{opt_problem_new} is non-concave. Additionally, the inequality conditions \eqref{limit_2_2} and \eqref{limit_2_3} are also non-concave due to the presence of $T_dT_s$ term in them. To deal with the non-concavity problem, we apply the second order Taylor expansion of the objective function and the \eqref{limit_2_2} and \eqref{limit_2_3} constraints as
\begin{align}\label{Taylor_expansion_1}
&\hat{f}^{(m+1)}\left(\mathbf{T}\right)= f\left(\mathbf{T}^{(m)}\right)+\nabla f\left(\mathbf{T}^{(m)}\right)^T\left(\mathbf{T}-\mathbf{T}^{(m)}\right)\nonumber\\
&+\frac{1}{2}\left(\mathbf{T}-\mathbf{T}^{(m)}\right)^T\left(\nabla^2 f\left(\mathbf{T}^{(m)}\right)\right)_+\left(\mathbf{T}-\mathbf{T}^{(m)}\right),
\end{align}
where $\mathbf{T}=\left[T_d \hspace{4mm} T_s\right]^T$, $f\in\{R_s,\rho^{(p_1,p_2)}_{s_a},\rho^{(p_1,p_2)}_{s_b}\}$, $\mathbf{T}_m$ is the point in which we approximate the $f$ in step $m$ of the algorithm and $(.)_+$ denotes the positive semi definite (PSD) part of the Hessian matrices. Therefore, we have successfully transfered the non-concave problem in \eqref{Opt_2} into a concave problem. Algorithm \ref{ALG_1} details the proposed SCP based algorithm, where $M$ is the maximum number of iterations.     
\begin{algorithm}
  \caption{Proposed SCP Based Iterative Optimization \label{ALG_1}}
  \textbf{Input}: $\mathbf{R}_{ar}, \mathbf{R}_{br}, \sigma_{ab}^2, P_a, P_b, \rho_{ab}, \rho_{ae}, \rho_{be}, \rho^t, T_p, M$\\
  \textbf{Output}: $T_d, T_s$
  \begin{algorithmic}[1] 
	\State \textbf{Initialization:} for $m=0$
\begin{itemize}
\item $T_d^{(0)}= 0.4 T_p$
\item $T_s^{(0)}= 0.16 T_p$
\end{itemize}
\Repeat{ (SCP Algorithm for $T_d$ and $T_s$)}
\State Update $m=m+1$.
\State Solve the concave program to obtain $T_d^{(m)}$ and $T_s^{(m)}$:
\begin{subequations}
\begin{gather}
\max_{T_d,T_s} \hspace{2mm}\hat{R}^{(m)}_s\left(\mathbf{T}\right) \nonumber
\end{gather}
\vspace{-5mm}
\begin{align}
&\hspace{-6mm}\textrm{s.t}: T_d+T_s\leq\frac{T_p}{2},\nonumber\\
& \hat{\rho}^{{(p_1,p_2)}^{(m)}}_{s_a}\left(\mathbf{T}\right)\le\rho^t,\nonumber\\
& \hat{\rho}^{{(p_1,p_2)}^{(m)}}_{s_b}\left(\mathbf{T}\right)\le\rho^t,\nonumber
\end{align}
\end{subequations}
\If{ $\rho^{{(p_1,p_2)}^{(m)}}_{s_a}\left(\mathbf{T}^{(m)}\right)\ge\rho^t$ \textbf{or} \\ \hspace{8.5mm} $\rho^{{(p_1,p_2)}^{(m)}}_{s_b}\left(\mathbf{T}^{(m)}\right)\ge\rho^t$}
\State $R_s^{(m)}\left(\mathbf{T}^{(m)}\right)=0$
\EndIf
\Until{$m=M$.} 
\State calculate $T_d^\star, T_s^\star$ as:
\begin{gather}
T_d^\star, T_s^\star = \argmax_{\mathbf{T}^{(m)}} \hspace{2mm}R_s^{(m)}\left(\mathbf{T}^{(m)}\right) \nonumber
\end{gather}
  \end{algorithmic}
\end{algorithm}

In Algorithm \ref{ALG_1}, due to the approximation of the two inequality constraints by the second order Taylor expansion, it is possible that the obtained $\mathbf{T}^{(m)}$ at each step does not satisfy the exact correlation constraints. Accordingly, we check this possibility in the following conditional expression to exclude the $\mathbf{T}^{(m)}$s which do not satisfy the correlation constraints from the final step. Moreover, it is not necessary to set feasible initial values for $T_d$ and $T_s$ as they are only used to approximate the following functions in a point. The algorithm will automatically converge to a feasible point and the non-feasible answers will be excluded through the conditional expression from the final results. Throughout this paper, we use the recommended initial values in all of the presented results. We further remark that computing the PSD part of the Hessian matrices in each iteration is not computationally demanding as the Hessians are $2\times2$ matrices. We further note that the proposed algorithm converges fast and as we will show in the numerical results section, it only requires few iterations to converge to an optimum point. 
\section{Numerical Results and Discussions}\label{numeric_discussion}
In this section, we intend to discuss the vital parameters affecting the SKG rate in the presence of a spatially correlated IRS and present our numerical results. In all the following results, unless otherwise stated, we assume an square IRS with $N_H=N_v=30$ and the element sizes of $d_H=d_V= \lambda/2$. The channel variances are given by $\sigma_{ij}^2= G_i+G_j+10\zeta_{ij}\log_{10}\left(d_{ij}/d_0\right)+\sigma_0^2$, where $i,j\in\{a,b,r,e\}$, $\sigma_0^2=-30$ dB is the path loss at $d_0=1$ m  and $G_i=G_j=4$ dBi denote the antenna gains at Alice, Bob and Eve and 0 dBi for Rose. Additionally, the distance between the nodes are considered as $d_{ab}=70$ m, $d_{ae}= 0.15$ m, $d_{be}=69.85$ m , $d_{ar}=4$ m, $d_{rb}=70.04$ m and $d_{re}=4$ m, where without loss of generality, we have assumed that the all nodes are located in a two dimensional plane and Eve is closer to Alice so that its observed channel can be correlated with the legitimate ones. Furthermore, the path loss exponents are assumed as $\zeta_{ab}=\zeta_{be}=4.8$, $\zeta_{ae}=\zeta_{ar}=\zeta_{er}=2.1$ and $\zeta_{br}=2.2$. Moreover, we assume the carrier frequency is $f_c= 1$ GHz, the system bandwidth is BW = 10 MHz and the noise figure at Alice, Bob and Eve to be NF = 5 dB.  

\begin{figure}
    \centering
	\includegraphics[width=3.5in,height=3.2in]{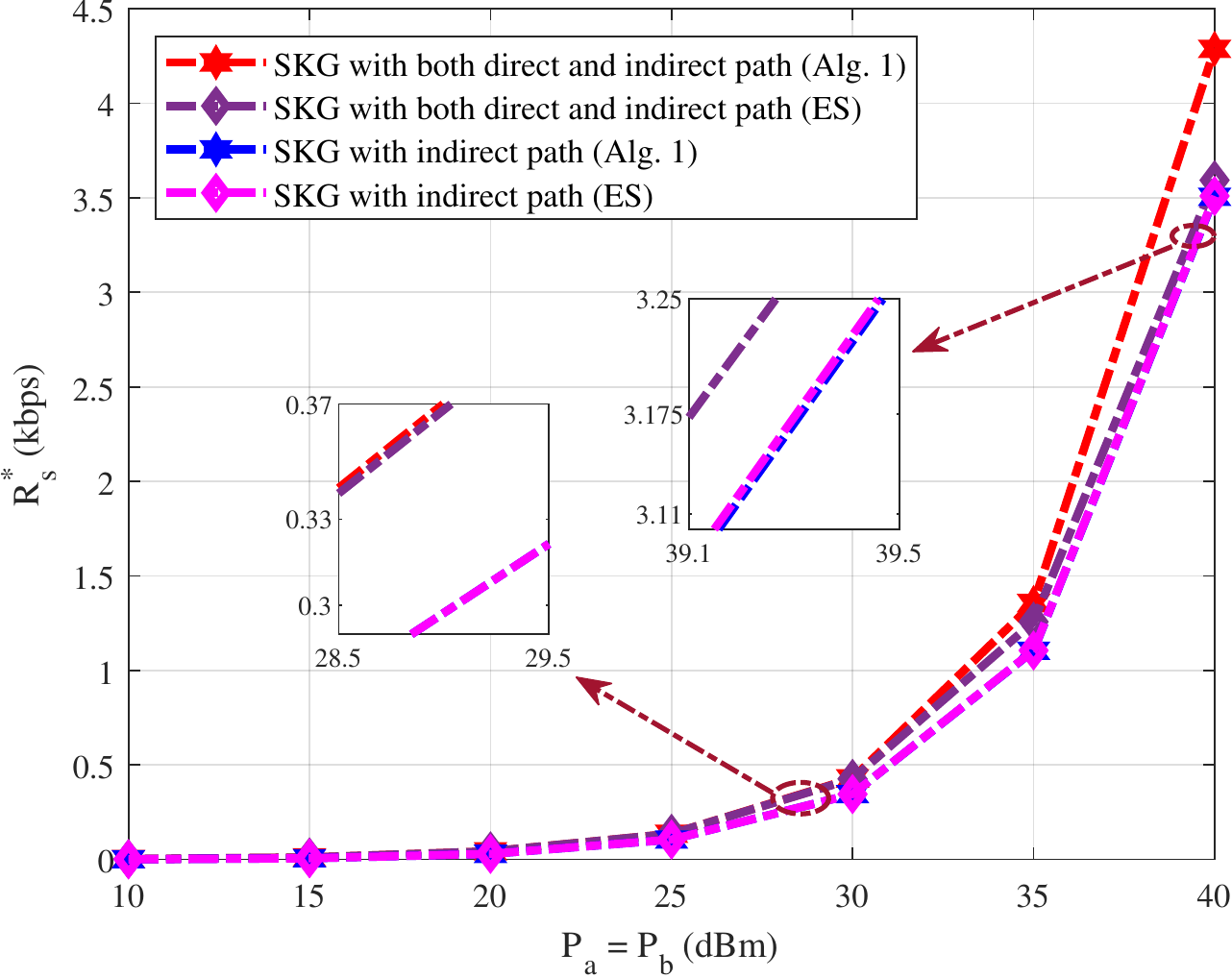}
	\caption{Contrasting the performance of Algorithm \ref{ALG_1} with the ES results for $M=20$.}\label{Optimization_Robustness}
\end{figure}

Fig. \ref{Optimization_Robustness}, contrasts the performance of our SCP based optimization algorithm with the results obtained by the exhaustive search (ES). We define $N_{T_d}=1/\zeta_{T_d}$ and $N_{T_s}=1/\zeta_{T_s}$, where $\zeta_{T_d}$ and $\zeta_{T_s}$ denote the search step size for $T_d$ and $T_s$ in our ES algorithm. Here, we have set $\zeta_{T_d}=\zeta_{T_s}= 10^{-2}$. Furthermore, we assumed the maximum permissible correlation between the channel samples $\rho^t=0.1$ which is shown to be sufficient in the SKG applications \cite{correlation_OFDM}. We have also accounted for the equal phase shift between the IRS elements in each probing. It can be observed that our algorithm outperforms the ES when direct and indirect probing samples are jointly considered in the SKG process. The gap between the two approaches becomes significant in high SNRs. Moreover, when we utilize only the indirect probing samples, the two algorithms lead nearly to the same rate. However, our algorithm is considerably time efficient comparing to the ES.     
\section{Conclusion}
In this contribution, for the first time we took into account the impact of spatial correlation between the IRS elements to study its impact on the SKG in a static environment. We showed that in comparison to the widely considered random independent phase shifts between the elements, equal phase shift can lead to higher SKG rates in most cases. We further proposed to exploit the randomness of direct and indirect paths separately to avoid strong correlation between the symbols of generated random key sequence. Moreover, we showed that it is vital to optimally allocate the direct and indirect time for probing the channel in each coherence time as the limited power can lead to significant CEE. Accordingly, we proposed an SCP based algorithm which is shown to be accurate and time efficient and was shown to significantly enhance the SKG rate.

\end{document}